\documentclass[conference]{IEEEtran}
\IEEEoverridecommandlockouts
\usepackage{preamble}

\begin{document}

\title{Authorization Propagation in Agent2Agent Systems: An Experimental Security Evaluation of Multi-Hop Delegation}

\author{\IEEEauthorblockN{Gabriele Donno, Kashif Mehmood, and Katina Kralevska} 
\IEEEauthorblockA{\textit{Department of Information Security and Communication Technology (IIK)} \\
\textit{NTNU\textemdash Norwegian University of Science and Technology, Trondheim, Norway}\\
Email: gabrido@stud.ntnu.no, \{kashif.mehmood, katinak\}@ntnu.no}
}

\maketitle

\begin{abstract}
Delegated Agent2Agent (A2A) workflows create a security problem when the agent sending a request is different from the principal from which the authority originated. This paper evaluates a layered downstream-verification design in a synthetic four-party healthcare reimbursement workflow. The prototype separates current-caller context, transitive authorization evidence, per-hop Ed25519 request authentication, and independent replay state for requests and authorization artifacts. Of 15 adversarial conditions, 14 are rejected; an isolated delegation-path test exposes the absence of semantic authority attenuation. A separate authorized-intermediary experiment shows that a holder of the shared authorization-evidence HMAC key can modify the represented authority and generate valid evidence. In the fixed-depth local benchmark, the full configuration has a median security-processing latency of 5.11 ms and a 3.87 kB serialized payload. The results show the value of separating request authentication from inherited authority while identifying origin-verifiable delegation and semantic attenuation as requirements for stronger A2A authorization propagation.
\end{abstract}

\section{Introduction}
Agentic AI workflows increasingly involve multiple independently implemented agents. The Agent2Agent (A2A) protocol provides a common model for agents to expose capabilities, exchange messages, and coordinate tasks~\cite{a2a_extensions}. In a delegated workflow, however, the agent sending the current request may be different from the participant from which the authority to perform the requested operation originated.

Consider a workflow \(A_0 \rightarrow A_1 \rightarrow A_2\). When \(A_2\) receives a request, it can authenticate \(A_1\) as the direct caller, but this alone does not show that the requested operation remains within the authority established earlier in the workflow. Conversely, valid propagated authorization evidence does not authenticate the request currently carrying it. A downstream participant therefore needs to verify both the current request and the authority inherited from previous hops.

A2A defines common structures for messages, tasks, discovery, and protocol extensions~\cite{a2a_extensions}. Each downstream interaction is a new A2A request, making the intermediary the direct caller of the next agent. We use the Secure Passport extension~\cite{secure_passport, habler2025} to carry current-caller context, while authorization is represented separately and propagated across the workflow. This separation allows the verifier to distinguish between who is making the current request and the authority the workflow carries.

\begin{table*}[t]
\centering
\caption{Security focus and trust assumptions of selected delegation mechanisms.}
\label{tab:related}
\small
\begin{tabular}{
    >{\raggedright\arraybackslash}p{2.1cm}
    >{\centering\arraybackslash}p{2.6cm}
    >{\centering\arraybackslash}p{2.6cm}
    >{\centering\arraybackslash}p{1.8cm}
    >{\centering\arraybackslash}p{1.7cm}
    >{\centering\arraybackslash}p{2.0cm}
    >{\centering\arraybackslash}p{2.5cm}
}
\toprule
\textbf{Approach} &
\textbf{Request authentication} &
\textbf{Origin authority} &
\textbf{Multi-hop delegation} &
\textbf{Replay controls} &
\textbf{Semantic Attenuation} &
\textbf{A2A evaluation}\\
\midrule
South \emph{et al.} \cite{south2025} & Partial & Yes & Partial & Partial & Policy-based & No \\
SAGA \cite{saga2026}                  & Yes     & No  & No      & Partial & No               & Yes \\
HDP \cite{hdp2026}                    & No      & Yes & Yes     & Yes     & App.-level& Partial \\
Progent \cite{progent2025}            & No      & No  & No      & No      & Yes              & No \\
AIP~\cite{aip2026}             & Bearer-token sig. & IBCT provenance & Yes & TTL-based & Yes & Binding specified \\
APS~\cite{aps2026} & Action-bound evidence & Signed principal binding & Yes & Nonce + ledger & Yes& No$^\dagger$ \\
\textbf{This work}                    & Yes     & Yes$^{*}$ & Yes$^{*}$ & Yes & \textbf{No} & Yes \\
\bottomrule
\multicolumn{7}{l}{\footnotesize $^{*}$Under the prototype's shared-HMAC trust assumptions; not origin-preserving against an authorized HMAC holder.} \\ \multicolumn{7}{l}{\footnotesize $^\dagger$APS does not specify a normative A2A binding; its released A2A adapter is experimental.}
\end{tabular}\vspace{-0.50cm}

\end{table*}

Related standards address complementary responsibilities: OAuth Token Exchange models delegation and impersonation~\cite{rfc8693}, RAR expresses structured authorization~\cite{rfc9396}, DPoP binds token use to key possession~\cite{rfc9449}, and HTTP Message Signatures authenticates selected message components~\cite{rfc9421}. We use these mechanisms as security reference points rather than claim implementation of their complete profiles.

Recent work has begun addressing these questions in agentic systems. \textit{South et al.} model authenticated, authorized, and auditable human-to-agent delegation~\cite{south2025}. \textit{SAGA} provides agent registration, authenticated discovery, lifecycle management, pairwise access control, and protected A2A communication~\cite{saga2026}. \textit{HDP} binds human authorization provenance to an append-only delegation sequence~\cite{hdp2026}, while \textit{Progent} addresses semantic privilege control and detects privilege expansion at tool-invocation boundaries~\cite{progent2025}. Broader work compares threats across emerging agent protocols~\cite{anbiaee2026}, while Agent Passport and Agent Identity proposals explore verifiable agent identity and delegation~\cite{aps2026, aip2026}. 

However, these mechanisms operate at complementary security boundaries. The narrower question investigated here is how downstream A2A participants can jointly verify the authorization inherited through a workflow, as well as the identity and integrity of the participant presenting the current request. Table~\ref{tab:related} positions the selected mechanisms according to signed per-hop request authentication, origin authority, multi-hop delegation, replay protection, semantic attenuation, and A2A evaluation. We study this question using a four-party A2A reimbursement workflow. The contributions of this paper are:
\begin{itemize}[leftmargin=*,nosep]
    \item \textbf{A security decomposition for delegated A2A workflows} that separates current-caller context, inherited authorization, per-hop request authentication, and independent replay domains.
    \item \textbf{A layered implementation} combining Secure Passport CallerContext, transitive authorization evidence, and signed per-hop requests without modifying the A2A core protocol.
    \item \textbf{An adversarial and local-overhead evaluation} across four incremental configurations, demonstrating both the protection gained and two residual design boundaries: lack of semantic scope attenuation and lack of origin preservation using a shared authorization key.
\end{itemize}

\section{System and Threat Model}
\label{sec:authorized}
\subsection{Workflow}
\begin{figure*}[t]
\centering
\begin{tikzpicture}[
    node distance=1.7cm and 2.9cm,
    every node/.style={font=\small},
    agent/.style={draw,rounded corners,minimum width=2.3cm,minimum height=.9cm,align=center},
    obj/.style={draw,dashed,rounded corners,align=left,inner sep=4pt},
    arr/.style={-{Latex[length=2mm]},thick}
]
\node[agent] (u) {Reimbursement\\Client $U$};
\node[agent,right=of u] (p) {Patient\\Agent $P$};
\node[agent,right=of p] (h) {Hospital\\Agent $H$};
\node[agent,right=of h] (i) {Insurance\\Agent $I$};

\draw[arr] (u) -- node[above]{start} (p);
\draw[arr] (p) -- node[above]{new A2A request} (h);
\draw[arr] (h) -- node[above]{new A2A request} (i);

\node[obj,below=.9cm of p] (pt) {
\textbf{Per-hop:}\\
$C_P,\;S_{PH}$
};
\node[obj,below=.9cm of h] (ht) {
\textbf{Per-hop:}\\
$C_H,\;S_{HI}$
};
\node[obj,below=.9cm of i] (trans) {
\textbf{Transitive:}\\
$A,\;E_A,\;D$
};

\draw[arr,dashed] (p) -- (pt);
\draw[arr,dashed] (h) -- (ht);
\draw[arr,dashed] (i) -- (trans);

\node[below=0.25cm of h,align=center,font=\small] {\hspace{-12.2cm}
$C_i$: current-caller context \quad \\ \hspace{-11.6cm}
$S_{ij}$: signed hop-specific request \quad \\ \hspace{-19.2cm}
$A$: structured authorization \quad \\ \hspace{-19.3cm}
$E_A$: authorization evidence \quad \\ \hspace{-20.5cm}
$D$: delegation path
};
\end{tikzpicture}
\caption{Four-party workflow separating hop-specific and transitive security objects.}
\vspace{-0.65cm}
\label{fig:workflow}
\end{figure*}
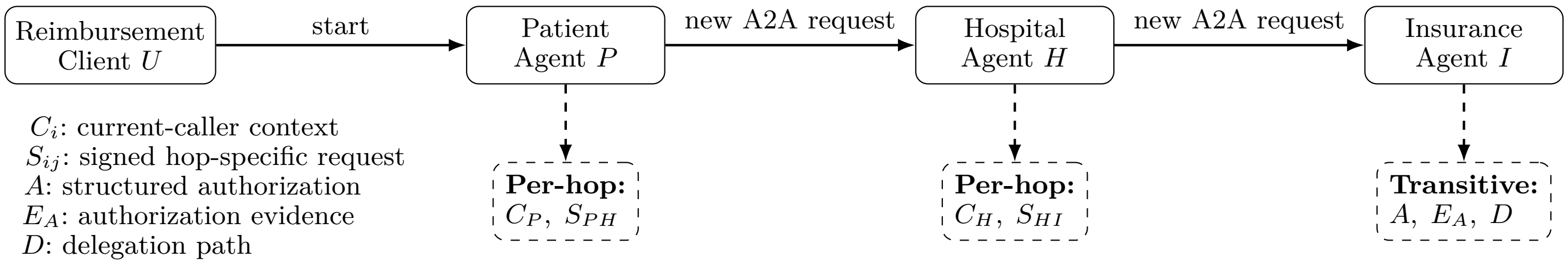

We evaluate a synthetic four-party healthcare reimbursement workflow (\figurename~\ref{fig:workflow}) chosen to separate caller identity, delegated authority, intermediate validation, and final decision-making. \(U\) is the Reimbursement Client, \(P\) the Patient Agent, \(H\) the Hospital Agent, and \(I\) the Insurance Agent. Healthcare data are synthetic and FHIR-inspired; full FHIR interoperability is outside the scope~\cite{fhir_r5}. The evaluated workflow uses a fixed four-party delegation path known to the participants; dynamic agent discovery, selection, and delegation-path construction are outside the scope of this experiment.

\subsection{Security Objects, Trust Assumptions, and Objectives}
Let \(C_i\) denote current-caller context, \(A\) the structured authorization, \(E_A\) its transitive evidence, \(D\) the represented delegation path, and \(S_{ij}\) the signed request from agent \(i\) to \(j\). In the prototype, the Reimbursement Client initiates the workflow request, while the Patient Agent constructs the structured authorization \(A\), creates the signed origin CallerContext, and issues the initial C2 authorization-evidence artifact. The artifact records an initial symbolic user-to-Patient delegation entry. C2 integrity uses a single shared prototype HMAC secret among components that create or verify authorization evidence. C1/C3 identities instead use statically provisioned Ed25519 public keys associated with the four participant identifiers. Key distribution, rotation, revocation, and an external PKI are outside scope.

\subsection{Adversary Model}
The adversarial matrix modifies selected security-sensitive fields while exercising the same production verification functions used during benign execution. Attacks are grouped into five families:
\begin{itemize}[leftmargin=*,nosep]
    \item \textbf{Sender attacks:} sender impersonation;
    \item \textbf{Request-integrity attacks:} request-text and protected-metadata
    modification;
    \item \textbf{Authorization attacks:} modification of maximum amount,
    beneficiary, coverage, treatment, or permitted action;
    \item \textbf{Delegation attacks:} unauthorized hop, destination substitution, and
    scope widening;
    \item \textbf{Freshness attacks:} request replay, authorization-artifact replay,
    and expiration.
\end{itemize}

A separate \emph{authorized-intermediary} case gives the Hospital Agent both its legitimate Ed25519 signing key and the shared HMAC secret used to protect authorization evidence. This stronger adversary models a participant that is trusted by the cryptographic construction but behaves maliciously with respect to the original authorization.
\section{Layered Authorization Verification}

\subsection{Current-Caller Context}
Configuration C1 adds Secure Passport v1 CallerContext. Its wire representation includes clientId, an optional sessionId, a signature, and structured state. Although the Secure Passport signature field is optional in the extension, the prototype requires it: the caller signs the canonicalized state using its Ed25519 identity, and the receiver verifies both the signature and that clientId matches the expected direct caller. Patient context includes identifiers for the patient, claim, and coverage, along with plan and consent state; Hospital context includes references to the claim, patient, and coverage, validation status, and the upstream caller identifier. Thus, $\textsf{VerifyContext}(C_i)\in\{\textsf{accept},\textsf{reject}\}$ authenticates the caller-specific context at the current hop but does not establish the workflow's transitive authority.

\subsection{Transitive Authorization Evidence}
Configuration C2 introduces a companion authorization evidence. At a conceptual level, it can be represented as
\begin{equation}
E_A = \left(id_o,\, H(C_o),\, H(A),\, D,\, t_{iat},\, t_{exp},\, n_A,\, v \right),
\end{equation}\vspace{-0.5cm}

where $id_o$ denotes the workflow origin identifier, $H(C_o)$ binds origin-related context, $H(A)$ binds the structured authorization details, $D$ records delegation evidence, $t_{iat}$ and $t_{exp}$ define validity, $n_A$ is an authorization-level nonce, and $v$ captures format/version information. The artifact is protected using
\begin{equation}
\tau_A = \operatorname{HMAC}_{K_A}(E_A).
\end{equation}\vspace{-0.5cm}

The implementation uses deterministic serialization before hashing and MAC calculation. HMAC therefore detects modification by parties outside the shared-key trust domain; Section V-B demonstrates why this construction does not preserve origin authorization against an authorized holder of $K_A$.

\subsubsection{Authorization Binding}
The verifier recomputes $h_A' = H(A')$ over the authorization details received in the current workflow and requires $h_A' = H(A)$ as stored in the protected evidence. Changes to the amount, beneficiary, coverage, treatment, or permitted action therefore produce a binding mismatch unless the attacker can also legitimately reconstruct the evidence artifact.

\subsubsection{Delegation Evidence}
The artifact carries an ordered representation of the delegation path. The Insurance Agent verifies the expected sequence of participants and the final destination. These checks detect unauthorized path substitution, but they are \emph{structural}. They do not determine whether authority is monotonically attenuated across delegation hops:
\begin{equation}
\textsf{VerifyPath}(D)= 1 \not\Rightarrow Scope_{i+1}\subseteq Scope_{i}.
\end{equation}\vspace{-0.5cm}

\subsubsection{Authorization Freshness and Replay}
Authorization validity is checked independently of replay state. An artifact must fall inside its validity interval and present an authorization nonce that has not been consumed: $ t_{iat}\leq t_{now}\leq t_{exp}$ and $n_A\notin R_A$. The authorization nonce is added to $R_A$ only after the relevant checks succeed.

$R_R$ and \(R_A\) are prototype replay-state sets maintained by the verifier. A forwarded A2A hop creates a new request and therefore a new \(n_R\); request and authorization replay are evaluated independently. An accepted nonce cannot subsequently be reused at the same replay state. The prototype does not evaluate atomic handling of concurrent duplicates, persistence across process restarts, or synchronization across replicated verifiers; consequently, no claims are made for those cases.

\subsection{Per-Hop Signed Request Authentication}
Configuration C3 adds a newly created signed envelope for each direct A2A request. The envelope conceptually contains
\begin{equation}
\begin{aligned}
S_{ij} = \bigl(&sender_i,\, receiver_j,\, msgID,\, t_R,\\
               &n_R,\, H(text),\, H(metadata) \bigr)
\end{aligned}
\end{equation}
and is authenticated using an Ed25519 signature
\begin{equation}
\sigma_i = \operatorname{Sign}_{sk_i}(S_{ij}),
\end{equation}
with verification
\begin{equation}
\operatorname{Verify}_{pk_i}(\sigma_i,S_{ij})=1.
\end{equation}
The protected-metadata digest covers the healthcare resources, authorization details, the current Secure Passport CallerContext, the transitive authorization-evidence artifact, the Hospital validation result, and the Hospital provenance. Other metadata fields are outside this digest unless explicitly added to the protected-field set. Ed25519 is used according to the standard EdDSA construction~\cite{rfc8032}. We refer to this project-specific mechanism as Signed A2A Request Authentication; it is not an implementation of HTTP Message Signatures. \textit{RFC~9421} instead serves as a standard for binding message components to a signer~\cite{rfc9421}. Request replay is distinct from authorization replay: $n_R \neq n_A, R_R \neq R_A$. A newly signed request can therefore be authentic while attempting to reuse authorization evidence that has already been consumed. In C3, the serialized CallerContext is additionally covered by the signed protected-metadata digest; consequently, modifications introduced after construction of the C3 envelope are rejected at the C3 layer before independent CallerContext processing is reached.

Request replay state is maintained independently by each receiving agent using a process-local nonce set. The request nonce is consumed after the C3 identity, freshness, digest, and signature checks succeed. Authorization replay uses a separate process-local cache: its nonce is checked during C2 verification and consumed by the Insurance Agent only after the associated authorization and application inputs have validated successfully. These caches are not persistent or distributed and are cleared on process restart. Concurrent duplicate handling, atomic consumption across replicas, and retry recovery are not evaluated.

\subsection{Fail-Closed Verification}
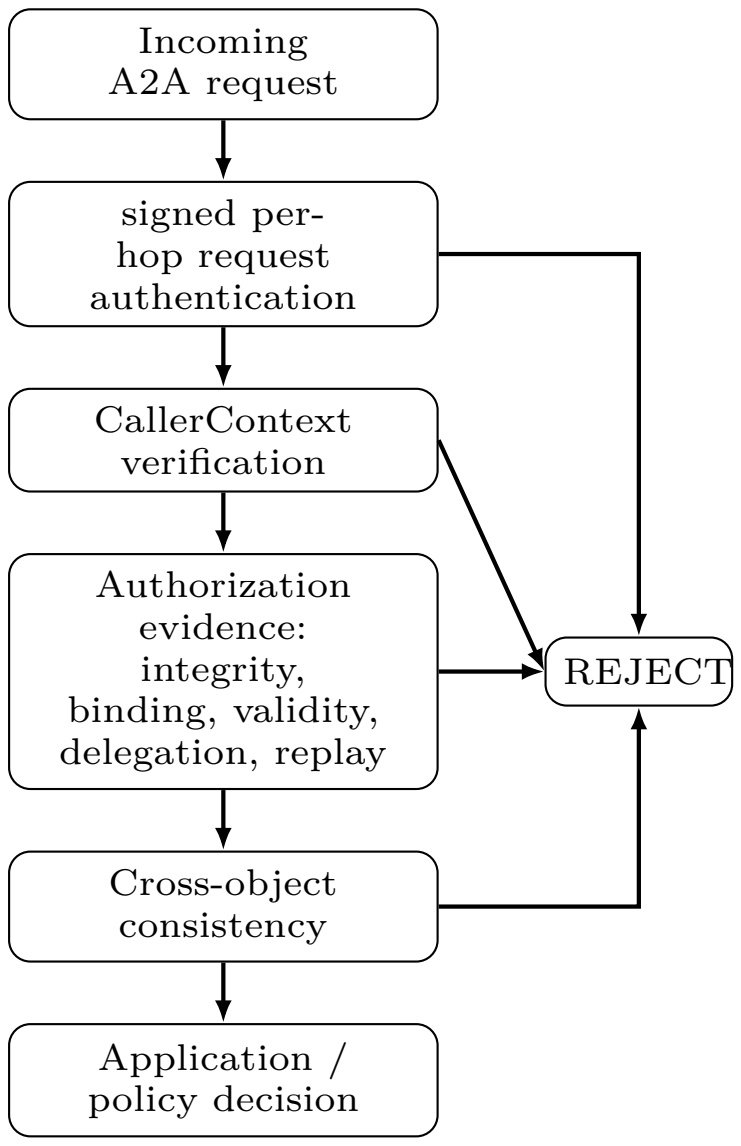
\begin{figure}[h]
\vspace{-0.4cm}
\centering
\begin{tikzpicture}[
    node distance=4.0mm,
    every node/.style={font=\scriptsize},
    box/.style={draw,rounded corners,align=center,text width=2.6cm,minimum height=.55cm},
    rej/.style={draw,rounded corners,align=center,text width=1.0cm,minimum height=.45cm},
    arr/.style={-{Latex[length=1.5mm]},thick}
]
\node[box] (in) {Incoming A2A request};
\node[box,below=of in] (r) {signed per-hop request authentication};
\node[box,below=of r] (c) {CallerContext verification};
\node[box,below=of c] (a) {Authorization evidence:\\integrity, binding, validity, delegation, replay};
\node[box,below=of a] (x) {Cross-object consistency};
\node[box,below=of x] (app) {Application / policy decision};
\node[rej,right=7mm of a] (rej) {REJECT};

\draw[arr] (in) -- (r);
\draw[arr] (r) -- (c);
\draw[arr] (c) -- (a);
\draw[arr] (a) -- (x);
\draw[arr] (x) -- (app);
\draw[arr] (r.east) -| (rej.north);
\draw[arr] (c.east) -- (rej.west);
\draw[arr] (a.east) -- (rej.west);
\draw[arr] (x.east) -| (rej.south);
\end{tikzpicture}
\caption{Fail-closed verification order; any failed mandatory check rejects the request.}
\vspace{-0.4cm}
\label{fig:verification}
\end{figure}

Cross-object consistency checks ensure that independently verified objects describe the same interaction: the authenticated request sender must match the current-caller identity, the request receiver and final delegation destination must identify the current verifier, and the received authorization must match the hash bound into \(E_A\). Thus, individually valid objects are rejected if their identities, destinations, or authorization bindings are inconsistent. Verification is performed before each receiving agent enters its application logic. A failed mandatory check prevents execution of the protected application operation and any downstream forwarding. \figurename~\ref{fig:verification} summarizes the order.

\begin{table*}[ht]
\centering
\caption{Adversarial outcomes and decisive rejecting layers.}
\label{tab:layered_attack_results}
\footnotesize
\setlength{\tabcolsep}{3.2pt}
\renewcommand{\arraystretch}{1.08}
\begin{tabular}{@{}p{4.25cm}p{1.35cm}p{7.2cm}p{2.25cm}p{2.25cm}@{}}
\toprule
\textbf{Attack / condition} &
\textbf{Result} &
\textbf{Decisive protection or reason} &
\textbf{Decisive layer} &
\textbf{Evidence} \\
\midrule

Sender impersonation &
Reject &
Ed25519 signature does not verify for the claimed sender &
C3 &
100/100 rejected \\

Modified request text &
Reject &
Signed request-text digest mismatch &
C3 &
100/100 rejected \\

Modified CallerContext &
Reject &
Protected-metadata digest mismatch &
C3 &
Functional test \\

Modified authorization-evidence metadata &
Reject &
Protected-metadata digest mismatch &
C3 &
100/100 rejected \\

Authorization-field tampering
(amount, beneficiary, coverage, treatment, action) &
Reject &
Authorization-details hash $H(A)$ no longer matches the protected binding &
C2 &
5/5 rejected \\

Unauthorized delegation hop &
Reject &
Delegation path contains an unexpected participant/destination &
C2 &
Functional test \\

Destination substitution &
Reject &
Final delegation entry does not terminate at the expected Insurance Agent &
C2 &
Functional test \\

Scope widening (isolated path test) &
\textbf{Accept} &
Added $approve_reimbursement$ to the Patient to Hospital scope; structural path verifier does not compare successive scopes &
C2 &
1/1 layer test \\

Request replay &
Reject &
Request nonce $n_R$ has already been consumed &
C3 &
100/100 rejected \\

Authorization-artifact replay &
Reject &
Authorization nonce $n_A$ has already been consumed &
C2 &
100/100 rejected \\

Expired authorization evidence &
Reject &
Authorization evidence falls outside its validity interval &
C2 &
Functional test \\

\bottomrule 
\end{tabular}\vspace{-0.30cm}

\end{table*}


\section{Prototype and Evaluation Methodology}

\subsection{Prototype Configurations}
The proof-of-concept was evaluated on a Windows 11 AMD64 host reporting an Intel64 Family 6 Model 140 Stepping 1 processor with eight logical CPUs, using Python 3.14.6, A2A SDK 1.0.0a0, and cryptography 46.0.5. The same four participants and reimbursement operation are used across all four incremental configurations; only the propagated security objects and corresponding verification logic differ. C0 implements the baseline A2A application workflow. C1 adds Secure Passport CallerContext verification. C2 adds the transitive authorization evidence, including authorization binding, delegation, validity, and replay checks. C3 adds Ed25519-authenticated per-hop request envelopes on top of C2.

\subsection{Functional Security Evaluation}
We first execute each configuration under benign conditions. The adversarial evaluation comprises fifteen scenarios reported in Table~\ref{tab:layered_attack_results}, together with one legitimate C3 control and a separate structural test confirming that request and authorization nonces belong to distinct replay domains. The authorized-intermediary experiment is evaluated separately because it assumes that the attacker legitimately possesses both a request-signing key and the shared authorization evidence key.

The repeated benchmark includes 100 benign executions per configuration, for a total of 400 benign runs. Five adversarial conditions—modified request text, modified protected metadata, sender impersonation, request replay, and authorization-artifact replay—are each repeated 100 times, giving 500 adversarial runs. These repetitions characterize deterministic behavior for the tested attack constructions and inputs; they are not estimates of detection probability for arbitrary attacks.

\begin{figure*}[t]
\centering

\begin{minipage}[t]{0.485\textwidth}
    \centering
    \includegraphics[width=\linewidth]{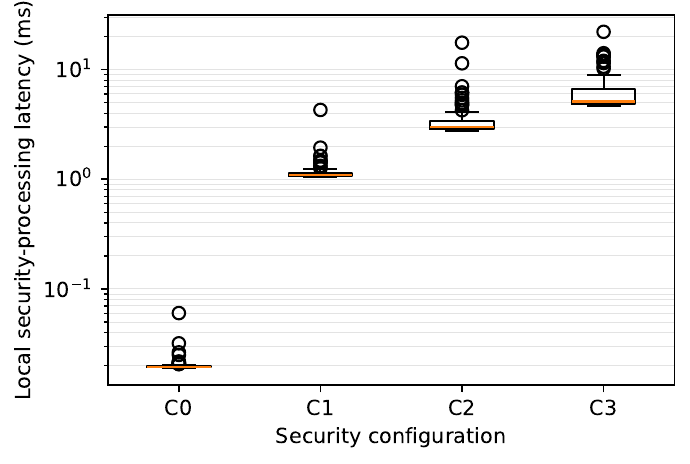}\\[-1mm]
    {\footnotesize (a) Benign local security-processing latency.}
\end{minipage}
\hfill
\begin{minipage}[t]{0.485\textwidth}
    \centering
    \includegraphics[width=\linewidth]{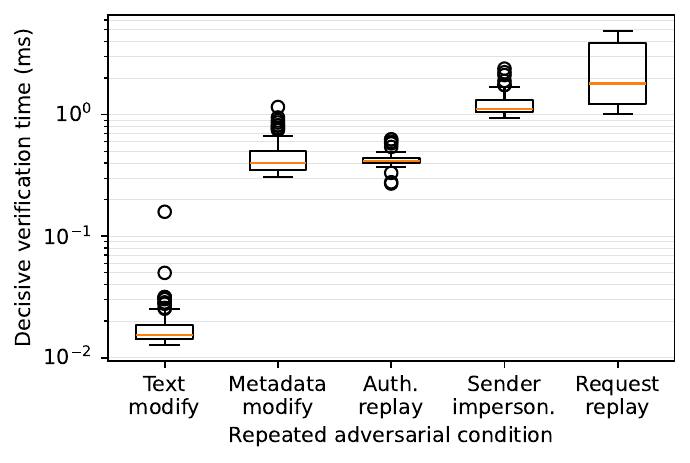}\\[-1mm]
    {\footnotesize (b) Verification time until rejection for repeated attacks.}
\end{minipage}

\caption{Local latency distributions over 100 repetitions per condition: (a) benign processing and (b) rejection time.}

\label{fig:latency_distributions}\vspace{-0.6cm}
\end{figure*}

For benign executions, timing starts immediately before local security verification and stops before application processing. For \figurename~\ref{fig:latency_distributions}(b), timing starts on entry to the verification routine containing the decisive check and stops when that check returns rejection. The corresponding configurations are C3 for request-text modification, protected-metadata modification, sender impersonation, and request replay, and C2 for authorization-artifact replay. Because these adversarial measurements begin at different verification routines, they characterize the local cost of the respective rejection paths and are not end-to-end cross-condition latency measurements. Network transport, HTTP/TLS framing, discovery, persistent storage, and cross-process communication are excluded.

\section{Experimental Results}
\label{sec:results}

\subsection{Layered Security Coverage and Fail-Closed Behavior}
\label{subsec:security_coverage}
All four configurations complete the benign workflow. The 15 adversarial scenarios in Table~\ref{tab:layered_attack_results} summarize both complete-request and isolated-layer adversarial tests. Fourteen of fifteen tested conditions are rejected at their intended verification boundary. The accepted case is an isolated delegation-path test in which $approve_reimbursement$ is appended to the Patient-to-Hospital scope. Because VerifyPath checks only the expected participant sequence and destination, the modified scope remains structurally valid, illustrating the limitation expressed in eq. 3. This demonstrates a missing semantic-attenuation check in the path verifier; it should not be interpreted as an end-to-end bypass by an attacker without the authorization-evidence key. Across the repeated benchmark, all 400 benign executions are accepted, and all 500 repeated adversarial executions are rejected. The replay tests additionally confirm that request and authorization nonces form independent replay domains.

\subsection{Trust-Boundary Experiments}
\label{subsec:trust_boundary_results}
The two accepted trust-boundary cases arise from different assumptions: scope widening exposes a lack of semantic attenuation, whereas the authorized-intermediary experiment exposes the shared-key trust model.
In this experiment, the Hospital Agent possesses both its legitimate Ed25519 signing key and the shared HMAC key ($K_A$). Starting from an authorization with a maximum amount of EUR 120, the Hospital changes it to EUR 9000 and recomputes the authorization-details hash. It then appends its legitimate Hospital-to-Insurance delegation entry; this operation recomputes the artifact HMAC using the shared C2 key. Finally, the Hospital constructs a fresh CallerContext and C3 request using its legitimate Ed25519 key. The resulting request passes C3 authentication, CallerContext verification, C2 integrity and origin-context verification, authorization binding, delegation-path verification, and the Insurance application checks. It characterizes precisely what the shared-key construction establishes:
\begin{align}
\mathrm{valid\ HMAC}
&\Rightarrow
\text{produced by a holder of } K_A,
\\
\mathrm{valid\ HMAC}
&\not\Rightarrow
\begin{aligned}
&\text{authorization unchanged from}\\
&\text{the workflow origin}.
\end{aligned}
\label{eq:hmac_origin_limit}
\end{align}
Thus, the HMAC establishes membership in the \(K_A\) trust domain, but not preservation of the originator's authority. The experiment therefore demonstrates a boundary of the shared-key design rather than a failure of HMAC integrity: every holder of \(K_A\) can cryptographically generate authorization evidence that is accepted as belonging to the same trust domain. Stronger designs can reduce this trust by preserving the origin's signature across the delegation chain, using per-hop signed delegation statements, or using attenuable capabilities in which an intermediary can only reduce authority. Macaroons~\cite{birgisson2014macaroons}, for example, support delegated authorization with restrictive caveats, while Lineage~\cite{stefanescu2026lineage} uses a resource-server-controlled delegation chain and enforces scope monotonicity. These alternatives provide stronger provenance or attenuation but introduce additional requirements for credentials, key management, or policy enforcement.

\begin{table*}[ht]
\centering
\caption{Incremental local-processing and serialized-payload costs.}
\label{tab:incremental_overhead}
\footnotesize
\setlength{\tabcolsep}{4.0pt}
\renewcommand{\arraystretch}{1.08}
\begin{tabular}{@{}p{0.75cm}p{5.05cm}p{2.85cm}p{1.45cm}p{2.15cm}p{1.75cm}p{2.05cm}@{}}
\toprule
\textbf{Cfg.} &
\textbf{Added mechanism} &
\textbf{Median (IQR) [ms]} &
\textbf{P95 [ms]} &
\textbf{$\Delta$ median [ms]} &
\textbf{Payload [B]} &
\textbf{$\Delta$ payload [B]} \\
\midrule

C0 & Baseline A2A/application workflow & 0.0195 (0.0005) & 0.0214 & -- & 513 & -- \\

C1 & Secure Passport CallerContext & 1.0883 (0.0626) & 1.4713 & +1.0688 & 996 & +483 \\

C2 & Transitive authorization evidence & 2.9986 (0.5278) & 5.8958 & +1.9103 &
2893 & +1897 \\

C3 & Per-hop signed request authentication & 5.1149 (1.7926) & 13.2127 & +2.1163 & 3870 & +977 \\
\bottomrule

\end{tabular}\vspace{-0.60cm}
\end{table*}

\subsection{Security--Performance Trade-off}
Table~\ref{tab:incremental_overhead} and \figurename~\ref{fig:latency_distributions} summarize the local cost of the incremental protection layers. C2 introduces the largest increase in serialized payload, adding 1897B for transitive authorization evidence. C3 adds a further 977B for per-hop request authentication, producing a complete serialized payload of 3870B. The full C3 configuration has a median local security-processing latency of 5.1149ms and a P95 of 13.2127ms, showing that the added protections increase both processing cost and execution-time variability. Separately recorded construction/signing and verification medians were 0.397/0.670 ms for C1, 0.591/1.608 ms for C2, and 1.413/2.803 ms for C3. Because several attacks target objects that do not exist in earlier configurations, adversarial conditions are evaluated at the first applicable security layer rather than being artificially replayed across all C0–C3 configurations; the table therefore distinguishes complete-hop tests from isolated-layer tests.

\figurename~\ref{fig:latency_distributions}(b) shows that median verification time until rejection ranges from 0.0155ms for request-text modification to 1.8038ms for request replay. Protected-metadata modification and authorization-evidence replay are rejected in approximately 0.4 ms, whereas sender impersonation reaches signature verification and requires 1.1079ms. Because the rejection measurements use condition-specific verification entry points, \figurename~\ref{fig:latency_distributions}(b) should be interpreted as the local cost of individual rejection paths rather than as a controlled latency comparison between attack classes. The benign measurements show the incremental local cost of adding authorization evidence and per-hop request authentication.
\section{Conclusion}
This paper separates inherited authorization from current-request authentication in a fixed-depth multi-hop A2A workflow. In the synthetic four-party experiment, the layered prototype rejects 14 of 15 adversarial scenarios. The full configuration has a median local security-processing latency of 5.11 ms and a 3.87 kB serialized payload; these measurements exclude network and distributed-system overhead. The isolated scope-widening case demonstrates the absence of semantic attenuation, while the authorized-intermediary experiment demonstrates the trust boundary of shared-HMAC authorization evidence. These results motivate origin-verifiable delegation and explicit authority attenuation in stronger A2A authorization designs. Dynamic delegation discovery, concurrent replay handling, restart persistence, and replicated verification remain outside the scope of the present evaluation.

\bibliographystyle{ieeetr}
\typeout{}
\bibliography{refs}

\end{document}